# Permeability and kinetic coefficients for mesoscale BCF surface step dynamics: discrete 2D deposition-diffusion equation analysis

Renjie Zhao,[1,2] James W. Evans[1,2,3,*], and Tiago J. Oliveira,[1,2,4]

[1]Ames Laboratory – USDOE, Iowa State University, Ames, Iowa 50011
[2]Department of Physics & Astronomy, Iowa State University, Ames, Iowa 50011
[3]Department of Mathematics, Iowa State University, Ames, Iowa 50011
[4]Departamento de Física, Universidade Federal de Viçosa, 36570-900, Viçosa, Minas Gerais, Brazil

*Corresponding author: evans@ameslab.gov

**ABSTRACT**

A discrete version of deposition-diffusion equations appropriate for description of step flow on a vicinal surface is analyzed for a two-dimensional grid of adsorption sites representing the stepped surface and explicitly incorporating kinks along the step edges. Model energetics and kinetics appropriately account for binding of adatoms at steps and kinks, distinct terrace and edge diffusion rates, and possibly asymmetric barriers for attachment to steps. Analysis of adatom attachment fluxes as well as limiting values of adatom densities at step edges for non-uniform deposition scenarios allows determination of both permeability and kinetic coefficients. Behavior of these quantities is assessed as a function of key system parameters including kink density, step attachment barriers, and the step edge diffusion rate.

## 1. INTRODUCTION

In 1951, Burton, Cabrera and Frank (BCF) introduced a strategy to describe the evolution of surface morphologies based upon coarse-graining of an atomistic-level treatment [1]. This BCF formulation applies for crystalline surfaces or epitaxial thin films with a well-defined terrace-step structure and with characteristic lateral lengths on the mesoscale. It describes step edges by continuous curves and uses appropriate evolution laws to track their motion [2,3]. This approach can be applied to analyze: step flow during deposition on vicinal surfaces; nucleation and growth of two-dimensional (2D) monolayer islands during deposition on flat surfaces where islands have a mesoscale lateral dimension due to facile surface diffusion; subsequent mound formation during unstable multilayer growth; and post-deposition island coarsening and mound decay [2,4,5].

The BCF prescription of step propagation is based on determination of fluxes of deposited atoms attaching to step edges. These fluxes in turn follow from analysis of the deposition-diffusion equations for the densities of diffusing adatoms on each of the terraces in a quasi-steady-state regime after imposing suitable boundary conditions (BC) at the step edges [1,2]. The original Dirichlet BC implemented by BCF treated steps as perfect traps for diffusing adatoms at which their density adopts its equilibrium value. This prescription was extended in 1961 by Chernov to incorporate possible inhibition in attachment of diffusing adatoms to steps, as characterized by kinetic coefficients, $K_{\pm}$, for ascending (+) and descending (-) steps [6]. Lower K values correspond to greater inhibition, so K measures the ease of attachment. There exist extensive analyses of step dynamics in both the diffusion-limited regime with high K, and the attachment-limited regime with low K [2,3].

In 1992, Ozdemir and Zangwill introduced the concept of step permeability or step transparency associated with direct transport across steps between terraces (without incorporation), the propensity for which is characterized by a permeability, P [7]. For P>0, deposition-diffusion equations on different terraces are coupled. Previous BCF type analyses incorporating step permeability have explored the transition to step flow [8] and step bunching or step instability on vicinal surfaces [3,9,10,11] during deposition. Other studies have explored the effect of permeability on second-layer nucleation during deposition [12,13], and on mound slope selection during unstable multilayer film growth [14]. In addition, the influence of permeability on post-deposition mound decay has been analyzed [15]. Analysis of experimental data has indicated significant permeability on Si(100) surfaces [16], but not on Si(111) surfaces [17]. In this study, our focus is not in BCF analyses of morphological evolution incorporating P, but rather on systematic derivation of P (and of the $K_{\pm}$) from an atomistic-level model. Furthermore, we will elucidate the dependence of P on key system parameters.

Next, we review theoretical formulations and analyses for $K_{\pm}$ and P. Some understanding has traditionally come from a steady-state analysis of discrete one-dimensional (1D) deposition-diffusion equation (DDE) models with the caveat that these models cannot account for step structure [4,5,18]. The classic analysis in the absence of permeability quantifies the decrease in K with increasing strength of an additional barrier for attachment to steps [19]. Permeability has also been incorporated into these 1D DDE models, at least in a simplistic fashion [18,20]. We will refine this treatment, thereby obtaining additional insight into the form of P. An important feature already apparent from these 1D analyses is the lack of a unique procedure to connect the discrete model to continuum formulation, and thereby to extract associated K's and P.

In this study, we improve substantially upon the above 1D treatments by implementing steady-state analyses based on appropriate discrete 2D deposition-diffusion equations (DDE) [18,21]. This formulation can directly incorporate basic features of step structure, in contrast to discrete 1D DDE's, thereby more realistically assessing $K_{\pm}$ and P. For uniform deposition on a perfect vicinal surface with a single type of step having symmetric attachment barriers (or no barriers), permeability does not affect behavior (see Sec.II), and one can directly determine the single K (but not P) [18]. However, if the single type of step has asymmetric attachment barriers, then one can obtain two relationships between the two distinct $K_{\pm}$ and P, but one cannot separately determine these quantities. Thus, to avoid this deficiency and enable determination of both K's and P, in the current study, we also explore behavior for non-uniform deposition on vicinal surfaces. Specifically, we allow different deposition rates on different terraces. Note that in the absence of the deposition on a specific terrace, a non-zero adatom density on that terrace can only develop due to step permeability. Just as for the discrete 1D DDE approach, there will be some non-uniqueness in the definition and extraction of $K_{\pm}$ and P from this formalism.

For comparison with our analysis based on discrete 2D DDE, it is appropriate to remark on other strategies for assessing P (and $K_\pm$) incorporating 2D surface geometries. A quite distinct approach is based on continuum models for step dynamics incorporating multiple density fields. Specifically, these approaches include separate density fields for edge and terrace adatoms, and also include additional information about step structure such as the mean kink density or even a kink density field [22,23,24]. Such analyses typically require some approximations, but they do provide expressions both for the kinetic coefficients, and for the step permeability, P. The predicted behavior for $K_\pm$ and P is intuitively reasonable. For example, P should decrease with increasing step attachment barrier, and with increasing step edge diffusivity allowing efficient transport to and incorporation at kink sites [22,23].

Lastly, we mention one targeted kinetic Monte Carlo simulation study of a stochastic lattice-gas model which assessed the propensity for step crossing as a function of step attachment barriers and kink separation [25]. Clearly, this crossing propensity is closely related to permeability, and these simulations indicated a qualitative dependence on model parameters consistent with previous studies.

In Sec.2, we review continuum and discrete DDE model formalisms, and present new results for a refined 1D DDE model. We present explicit results for $K_\pm$ and P from extensive numerical analysis of the discrete 2D DDE for symmetric (or zero) attachment barriers in Sec.3, and for asymmetric attachment barriers in Sec.4. Application of the results to specific systems is discussed, and conclusions are presented in Sec.5.

## 2. DISCRETE DEPOSITION-DIFFUSION EQUATION (DDE) FORMALISMS

First in Sec.2A, to provide further background on the BCF formulation, we briefly review the continuum BCF formulation. This formulation considers the adatom density per unit area, $\rho(\underline{x}, t)$, at lateral position $\underline{x} = (x, y)$, where in our analysis steps on the vicinal surface will be aligned with the x-direction. The kinetic coefficients, $K_\pm$, for attachment to ascending (+) and descending (-) steps and the permeability, P, appear in the boundary conditions to the continuum deposition-diffusion equations for $\rho(\underline{x}, t)$. The basic model includes uniform deposition rate per unit area, $\mathcal{F}$. Refined models will include different deposition rates $\mathcal{F}_{t1}$ and $\mathcal{F}_{t2}$ on alternating terraces. Another key parameter is the terrace diffusion coefficient, D.

Next, we review discrete 1D DDE formulation in Sec.2B and 2C, and the 2D DDE formulation in Sec. 2D and 2E, as well as the procedure for obtaining $K_\pm$ and P. The 1D formulation will correspond to a reduced version of the 2D formulation. Thus, we here highlight some basic features of the discrete 2D DDE formulation. Adatoms reside at a square array of adsorption sites labeled (i, j) with lattice constant 'a' and with steps on the vicinal surface aligned with the i-axis. The adatom density at these sites is denoted by n(i,j), leaving implicit the t-dependence, and corresponds to the probability that site (i, j) is occupied. One has n(i,j) << 1 under typical conditions. The basic model includes uniform deposition at rate F per site, and refined models include different rates $F_{t1}$ and $F_{t2}$ on alternating terraces. Terrace diffusion corresponds to hopping to nearest-neighbor (NN) empty sites at rate h per direction. Diffusive dynamics at step edges differs from that on terraces, we comment on some key aspects below.

In general, we include possibly asymmetric step attachment barriers leading to reduced rates $h_\pm$ for hopping to a straight step edge relative to the terrace hop rate, h. We will write $h_\pm = \exp(-\beta\delta_\pm)h$, where $\delta_\pm$ denote additional attachment barriers, assuming a common prefactor for all hops. Here + (-) corresponds to attachment to ascending (descending) steps, and we set $\beta = 1/(k_B T)$ where T is the surface temperature and $k_B$ is the Boltzmann's constant. The extra barrier $\delta_+$ to attach to a descending step is described as the Ehrlich-Schwoebel barrier. Processes occurring at the step edge such as detachment, edge diffusion, equilibration or incorporation, will be described in more detail below.

In our discrete DDE models, we will also allow the possibility of anisotropic NN lateral interactions mimicking, e.g., a fcc(110) versus a fcc(100) surface. However, for simplicity, we will retain isotropic terrace diffusion. NN attractions in the direction orthogonal (parallel) to the step will have strength $\phi_\perp > 0$ ($\phi_\parallel > 0$). Consequently, step edge adatoms are bonded to the straight step by a NN attraction of strength $\phi_\perp > 0$ orthogonal to the step, and the rates for detachment from straight steps are given by $\exp(-\beta\phi_\perp)h_\pm$ according to detailed-balance. Adatoms at kink sites have an additional bond of strength $\phi_\parallel > 0$ parallel to the step, and thus a total bonding of $\phi_b = \phi_\perp + \phi_\parallel$. Naturally setting the density of atoms at kink sites as n(kink) = 1, it follows that the equilibrium densities in the absence of deposition are

$n_{eq}(\text{edge}) = \exp(-\beta\phi_\parallel)n(\text{kink}) = \exp(-\beta\phi_\parallel)$, and  (1)

$n_{eq}(\text{terrace}) = \exp(-\beta\phi_\perp)n_{eq}(\text{edge}) = \exp(-\beta\phi_b)n(\text{kink}) = \exp(-\beta\phi_b) \equiv n_{eq}$,  (2)

for step edge and terrace adatoms, respectively.

In <u>coarse-graining</u> from discrete to continuous models, it is natural to make the correspondence $\underline{x} = (ai, aj)$ for the lateral position, and $\rho(\underline{x}, t) = a^{-2} n(i,j)$ for the adatom density. More specifically, the rescaled spatially continuous density field $a^2\rho(\underline{x}, t)$ should be regarded as passing smoothly through the discrete densities, n(i,j). Thus, one has that the continuum equilibrium density satisfies $\rho_{eq} = a^{-2}n_{eq}$, with $n_{eq}$ defined by (2). Sometimes it will be instructive to introduce "excess densities" $\delta\rho = \rho - \rho_{eq}$ or equivalently $\delta n(i,j) = n(i,j) - n_{eq}$. In this mapping between discrete and continuous models, the deposition rate per unit area in the continuum model is given by $\mathcal{F} = a^{-2} F$, or $\mathcal{F}_{ti} = a^{-2} F_{ti}$, and the terrace diffusion rate satisfies $D = a^2 h$. Below, we shall see that there is some ambiguity or flexibility in mapping discrete onto continuous models which result in slightly different expressions for $K_\pm$ and P. Also, in the discrete model, there is potentially a significant contribution to step propagation velocity from adatoms depositing directly at the step edge (i.e., at the row of sites directly adjacent to the step edge) [21]. The precise form of this contribution is tied to the definition of $K_\pm$. Such a contribution is clearly absent in the continuum formulation.

## 2A. REVIEW OF CONTINUUM BCF FORMULATION

The traditional continuum BCF formulation considers uniform deposition where again $\mathcal{F}$ denotes the deposition flux per unit area, and D is the terrace diffusion coefficient. One performs a quasi-steady-state analysis of the continuum deposition-diffusion equation for the adatom density, $\rho(\underline{x}, t)$, of the form

$$\partial/\partial t\, \rho(\underline{x}, t) = \mathcal{F} + D\nabla^2 \rho(\underline{x}, t) \approx 0. \qquad (3)$$

General Chernov-type BC's at permeable step edges have the form

$$J_\pm = \pm D\, \nabla_n \rho|_\pm = J_{K\pm} + J_P, \text{ where } J_{K\pm} = K_\pm(\rho_\pm - \rho_{eq}) \text{ and } J_P = P(\rho_\pm - \rho_\mp). \qquad (4)$$

Here, $J_\pm$ denote the net diffusion fluxes for attachment to an ascending step from the terrace below (+), and to a descending step from the terrace above (-), respectively. $J_\pm$ incorporate two types of contributions. $J_{K\pm}$ denote those associated with step attachment and detachment, where $K_\pm$ are the corresponding Chernov kinetic coefficients. $J_P$ denotes the contribution from step crossing, where P is the step permeability. $\nabla_n$ denotes the gradient normal to the step. $\rho_\pm$ are the limiting values of the terrace adatom density approaching the step on the lower (+) and upper (-) terrace, respectively, and $\rho_{eq}$ denotes the equilibrium adatom density at the step. See Fig.1. It is common to set $K_\pm = D/\Gamma_\pm$ where $\Gamma_\pm$ denote the attachment lengths, and $P = D/\Gamma_P$ where $\Gamma_P$ denotes the permeability length. Then, large $\Gamma$'s implying difficult attachment or step crossing. The sign convention is chosen for a vicinal surface descending to the right and where we define net attachment fluxes to be positive, $J_\pm > 0$.

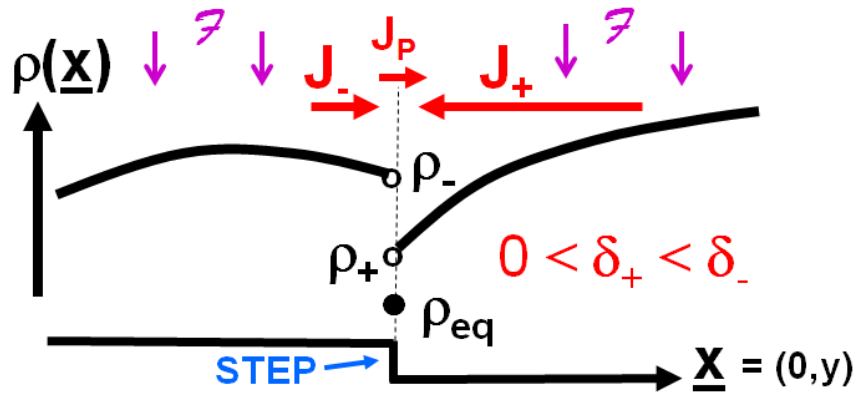

Fig.1. 1D schematic of adatom density and diffusion flux behavior at a step with asymmetric attachment. The total diffusion fluxes $J_+ = K_+(\rho_+ - \rho_{eq}) + P(\rho_+ - \rho_-)$ and $J_- = K_-(\rho_- - \rho_{eq}) + P(\rho_- - \rho_+)$ reaching the ascending and descending steps, respectively, and the flux across the step due to permeability $J_P = P(\rho_- - \rho_+)$ are indicated.

Next, we discuss one example of behavior for a non-uniform deposition flux, specifically for flux switching between a larger value, $\mathcal{F}_{t1}$, and smaller value, $\mathcal{F}_{t2}$, on alternating terraces, and where there is a symmetric attachment barrier so that $K_+ = K_- = K$. As illustrated in Fig.2, the density profile has mirror symmetry about the center of each terrace, so the attachment flux has the same value of both sides of each terrace. We can readily extract some basic information about the density profile by balancing deposition and attachment fluxes on each terrace, i.e., $J_i = ½ \mathcal{F}_{ti} W$. Let $\rho_i$ denote the adatom density at the edge of terrace $i = 1$ or $2$, and $\delta\rho_i = \rho_i - \rho_{eq}$ denote the corresponding excess density. Then, by adding and subtracting the expressions for $J_1$ and $J_2$ and rearranging the results, we obtain the following relations

$$\delta\rho_1 - \delta\rho_2 = ½ (\mathcal{F}_{t1} - \mathcal{F}_{t2})W/(K+2P) \text{ and } \delta\rho_1 + \delta\rho_2 = ½ (\mathcal{F}_{t1} + \mathcal{F}_{t2})W/K. \tag{5}$$

A particularly instructive case is when $\mathcal{F}_{t2} = 0$, so then the adatom density is uniform on terrace 2, and the associated excess density satisfies

$$\delta\rho_2 = ½ W \mathcal{F}_{t1} P/[K(K+2P)]. \tag{6}$$

Thus, the excess density on terrace 2 in the absence of deposition is only non-zero if $P > 0$, and its magnitude provides a measure of P.

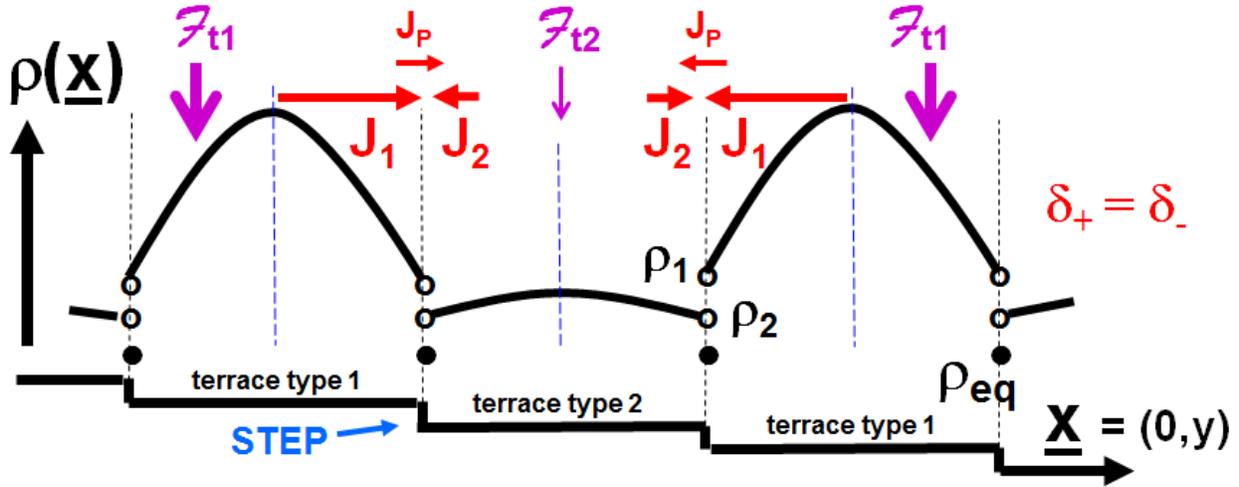

Fig.2. 1D schematic of adatom density and diffusion flux behavior on a vicinal surface where the flux has different values ($\mathcal{F}_{t1} > \mathcal{F}_{t2}$) on alternating terraces, and with a symmetric step attachment barrier.

### 2B. DISCRETE 1D DDE MODEL: BASIC FORMULATION

For a vicinal surface with straight parallel steps aligned with the i-axis (i.e., the x-direction), the simplest picture anticipates that the adatom density is independent of position along the step, but varies across the terrace. Thus, $n(i,j) = n(j)$ depends only on the label j of the rows of sites parallel to the step and, equivalently, $\rho(\underline{x}, t) = \rho(y, t)$ in the continuum formulation. This feature reduces the discrete 2D DDE formulation to a discrete 1D DDE formulation. The specific form of the discrete 1D DDE for the adatom density at rows of sites away from the step edges is

$$d/dt\, n(j) = F + h\, \Delta_j n(j) \text{ where } \Delta_j n(j) = n(j+1) - 2n(j) + n(j-1), \tag{7}$$

where again h is the terrace hop rate. Separate equations are needed for the adatom density at or adjacent to steps, where the rates for hopping might be impacted by step attachment barriers, and by distinct processes at the step edge. See Appendix A.

A detailed schematic of behavior in our discrete 1D DDE model is provided in Fig.3. In this prescription, n(j≠0) denote the densities of terrace adatoms, and n(j=0) denotes the density of step edge adatoms. As indicated in the introduction to Sec.2, our model includes reduced attachment rates $h_\pm$ for hopping to the step edge (j=0) relative to h, and detachment rates $\exp(-\beta\phi_\perp)h_\pm$ reflecting bonding of edge adatoms to the step with attraction $\phi_\perp$. In addition, our model incorporates the feature that adatoms which hop to the step edge are not immediately incorporated into the growing crystal. This, in turn, reflects the feature that in realistic 2D models, adatom incorporation effectively only occurs at kink sites which can be rare on close-packed steps. The rate of incorporation or equilibration in our 1D model is denoted by an additional parameter, $\nu$, defined through the relation

$$d/dt\, n(0)|_{relax} = -\nu\, [n(0) - n_{eq}(0)]. \tag{8}$$

Below, we will introduce a naturally rescaled relaxation rate, r, and relate r and $\nu$ to permeability. We note that our model is similar in spirit, but different in detail from a model of T. Zhao *et al.* [20].

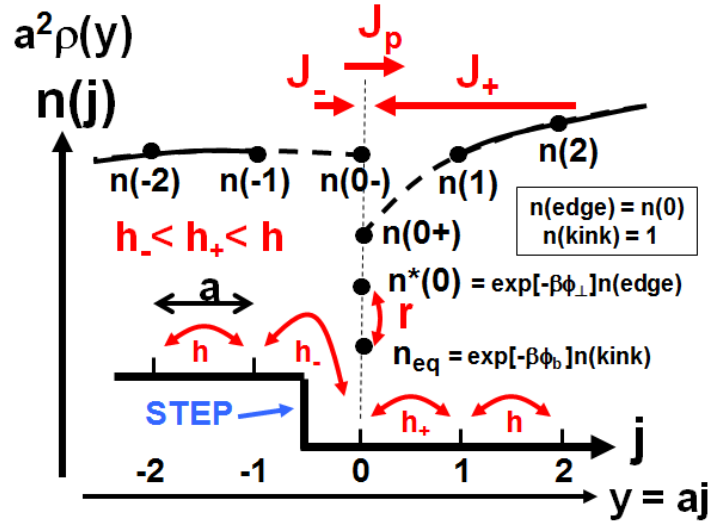

Fig.3. Schematic of our discrete 1D DDE model. n(j) denotes the adatom density on row j of sites, where j=0 corresponds to the step edge. Note that step detachment contributions are $d/dt\, n(\pm 1)|_{detach} = \exp(-\beta\phi_\perp)h_\pm\, n(0) = h_\pm\, n^*(0)$. Note that $n^*(0)$ coincides with n(0+) [n(0-)] in the case of zero attachment barrier $\delta_+$ [$\delta_-$].

Given the feature that our discrete 1D DDE model should mimic the more realistic 2D DDE model, the equilibrium edge adatom density $n_{eq}(0) = n_{eq}(edge)$ in the 1D model should be enhanced by a factor of $\exp(\beta\phi_\perp)$ relative to the equilibrium terrace density, $n_{eq} = \exp(-\beta\phi_b)$. Similar enhancement should persist in the presence of deposition. This suggests that the edge atom density n(0) = n(edge) is naturally rescaled to n*(0) = n*(edge) = exp(-βϕ⊥)n(0) making it comparable in magnitude to terrace densities. Using this notation, one has d/dt n(0)|relax = -r [n*(0) - neq] with the rescaled rate, r = exp(+βϕ⊥)ν.

## 2C. DISCRETE 1D DDE MODEL: EXTRACTION OF $K_\pm$ AND P

Analysis involves obtaining expressions relating diffusion fluxes to the step edge, $J_\pm$, and adatom densities "at" the step edges, $\rho_\pm$, and from these extract $K_\pm$ and P. However, there is some flexibility in the identification of both $J_\pm$ and $\rho_\pm$. The 1D DDE diffusion fluxes are most naturally identified from sites j = ±1 to the step edge. Significantly, results depend upon the identification of $\rho_\pm$. These could most simply be taken as the densities, $a^{-2}$ n(±1), at sites adjacent to the step site j=0 (labeled as the "non-extrapolation" case N). Alternatively, they can be chosen as $a^{-2}$ n(0±), where n(0±) are obtained by suitably "analytically extending" terrace adatom densities, n(j), to the site j = 0 (labeled as case E for "extrapolate" or "extend"). We also note the non-trivial result for uniform deposition that the rescaled density at the step edge, n*(0), corresponds to the extrapolated density, n(0+) [n(0-)], in the case of no additional attachment barrier $\delta_+$ [$\delta_-$]. Different choices produce slightly different results for $K_\pm$ and P. We note that the same applies for other formulations which are also possible, e.g., determining the fluxes between sites j = ±2 and j = ±1, a choice denoted by M. See Appendix A for further discussion.

Here, we just report the results of the above analysis for cases N and E. For case N, where $\rho_\pm$ are identified as $a^{-2}$ n(±1), one obtains

$$K_\pm(N) = arh_\pm/(h_+ + h_- + r), \text{ so } \Gamma_\pm(N) = ah(h_+ + h_- + r)/(rh_\pm), \text{ and} \qquad (9)$$

$$P(N) = ah_+h_-/(h_+ + h_- + r), \text{ so } \Gamma_P(N) = ah(h_+ + h_- + r)/(h_+ h_-). \qquad (10)$$

For the case E, where $\rho_\pm$ are identified as the analytically extended $a^{-2}$ n(±0), one obtains

$$K_\pm(E) = ar(h/h_\pm - 1)^{-1}/[(h/h_+ - 1)^{-1} + (h/h_- - 1)^{-1} + (r/h)], \text{ and} \qquad (11)$$

$$P(E) = ah(h/h_+ - 1)^{-1}(h/h_- - 1)^{-1}/[(h/h_+ - 1)^{-1} + (h/h_- - 1)^{-1} + (r/h)], \qquad (12)$$

from which one can obtain corresponding expressions for $\Gamma_\pm(E)$ and $\Gamma_P(E)$.

Next, we discuss behavior in key limiting regimes, and also compare results of the E and N treatments. In the limit of instantaneous incorporation or equilibration at step edges, r→∞, one obtains $K_\pm(N) \to ah_\pm$ so $\Gamma_\pm(N) \to ah/h_\pm$, and $K_\pm(E) \to ah/(h/h_\pm -1)$ so $\Gamma_\pm(E) \to a(h/h_\pm -1)$. One also naturally obtains P(N) → 0, so $\Gamma_P(N) \to \infty$, and P(E) → 0 so $\Gamma_P(E) \to \infty$. In the limit of vanishing attachment barriers, $h_+ = h_- \to h$, one obtains $\Gamma_\pm(N) \to a(2 + r/h)/(r/h)$ and $\Gamma_P(N) \to a(2 + r/h)$, versus $\Gamma_\pm(E) \sim 2a/(r/h)$ and $\Gamma_P(E) \to 0$. Thus, for general $h_\pm \neq h$, we find that values of $\Gamma_\pm(N)$ and $\Gamma_\pm(E)$ differing by 'a' as r → ∞. Similarly, for $h_\pm = h$, we find that $\Gamma_\pm(N) \to a$ but $\Gamma_\pm(E) \to 0$, as r/h → ∞. In both cases, values of $\Gamma_\pm(N)$ and $\Gamma_\pm(E)$, or difference between them, are far below the terrace width,

W, and so solution of the appropriate boundary value problem continuum deposition-diffusion equations will produce similar behavior. Finally, we note that in the regime of strong inhibition of attachment to steps $h_\pm << h$ or $\beta\delta_\pm >> 1$, expressions from both formulations agree and reduce to $K_\pm \approx ah \exp(-\beta\delta_\pm)$ and $P \approx ah \exp(-\beta\delta_+ -\beta\delta_-)/(r/h)$.

We note some similarity between the forms in (11) and (12), and the corresponding expressions in T. Zhao *et al.* [20] who also applied a discrete 1D DDE approach which incorporated extrapolation of terrace densities to step edges. In particular, the combinations $(h/h_\pm - 1)$ naturally appear as a result of analytic extension or extrapolation. However, T. Zhao *et al.* introduce a probability, $p_{inc}$, for incorporation of any atom reaching a step. There is not a simple mapping between r and $p_{inc}$, although one has $p_{inc} \to 1$ (0), as $r/h \to \infty$ (0) [26]. It is also appropriate to note that (9) and (10) match the corresponding expressions obtained by Pierre-Louis [22] from quite different continuum approach with multiple diffusion fields.

An additional question of particular relevance is how $\nu$ or r, and thus P and K's, are related to additional physical parameters in a realistic 2D model, such as the edge diffusion rate and kink density. Such parameters do not appear explicitly in our 1D model. Consider systems with <u>facile edge diffusion</u> associated with hop rate $h_e > h$, and steps with a mean kink separation $L_k = l_k a$. Then, the characteristic time for edge diffusion mediated incorporation of a mobile edge atom at a kink should scale like $\tau \sim (l_k)^2/h_e$ based on the Einstein diffusion relation. The relaxation rate, $\nu$, should then be given by $\nu = 1/\tau$ so that $r \sim \exp(+\beta\phi_\perp)h_e/(l_k)^2$. This form for $\nu$ was suggested previously in Ref.[22]. It produces the scaling $K_\pm \sim (ah)/(l_k)^2$, as $l_k \to \infty$. Note this latter scaling applies more generally than in the case of facile edge diffusion as demonstrated previously from analysis of the discrete 2D DDE for symmetric attachment barriers, $h_+ = h_-$ [21].

Finally, we mention that conventional versions of discrete 1D DDE models set $r = \infty$ corresponding to instantaneous equilibration or incorporation at the step edge [4,5]. Then, permeability can still be incorporated (somewhat artificially) by modifying the model to introduce direct hopping across steps, e.g., between sites $j = 1$ and $j = -1$ at rate $h_p$. See Appendix A for the associated $K_\pm$ and P.

## 2D. DISCRETE 2D DDE MODEL: FORMULATION, DEFINITION OF $K_\pm$ AND P

Figure 4 shows a schematic of our discrete 2D DDE model. We consider a perfect vicinal surface with straight parallel steps aligned with the i-direction and separated by terraces of equal width $W = w\,a$ (for integer w). We identify rows $j = 0$, $j = \pm w$, $j = \pm 2w$, etc., as step edge rows, and rows $0 \leq j < w$, $w \leq j < 2w$, etc. as being on the same terrace. The vicinal surface descends with increasing j. Kink sites are specified to be periodically distributed along step edge rows with separation $L_k = l_k a$, so kinks are located at $(i, j) = (nl_k, mw)$ for integer n and m. Thus, all steps are equivalent. The discrete 2D DDE model accounts for variation of the adatom density, $n(i, j)$ at site $(i, j)$, both along the steps as well as across the terraces. Away from the step edges, the 2D DDE have the form

$$d/dt\, n(i,j) = F + h\, \Delta_{i,j}\, n(i,j), \tag{13}$$

where $\Delta_{i,j}$ is the discrete 2D Laplacian, so that $\Delta_{i,j} n(i,j) = n(i+1,j) + n(i,j+1) + n(i-1,j) + n(i,j-1) - 4n(i,j)$. As discussed further below, refined equations are needed at and adjacent to step edges and at kink sites where the hop rates are modified reflecting possible attachment barriers at steps and binding at step edges. Kink sites constitute both a source and a sink for adatoms and the density of adatoms at kink sites is set to unity (as described above).

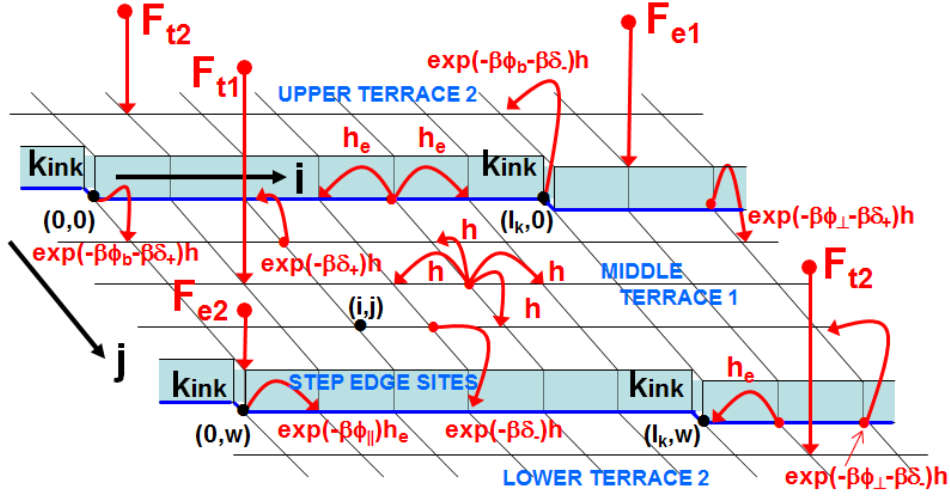

Fig.4. Schematic of our discrete 2D DDE model, indicating the rates for various hopping processes and depositions. Solution of the DDE equations just require analysis of densities in a periodic unit cell of sites, $0 \leq l < l_k$ and $0 \leq j < 2w$. The unit cell consists of a strip between adjacent kink sites spanning two adjacent terraces.

In our model with nearest-neighbor (NN) adatom attractions, the interaction $\phi_{||}$ parallel to the steps controls the kink separation, $L_k \approx \frac{1}{2} a \exp(\frac{1}{2}\beta\phi_{||})$ [2]. Model behavior will depend on $\phi_{||}$ only through kink separation as demonstrated in Ref.[18] by analysis of the discrete 2D DDE's for suitably rescaled adatom densities at step edges and kink sites. Model behavior also depends on the NN interaction $\phi_\perp$ controlling edge adatom bonding to the step and the rate, $h_e$, for diffusion along a straight step. Interestingly, from an analysis of rescaled 2D DDEs, it is possible to show that these parameters always appear in the combination $\exp(\beta\phi_\perp)h_e/h$. Therefore, selecting $h_e = \exp(-\beta\phi_\perp)h$ makes the model independent of $\phi_\perp$. In some sense, the steps are "invisible" for this choice, since the effect of binding to the steps is compensated for by the effect of slower edge diffusion than terrace diffusion. We use this choice for most results presented in the following sections. In the general case with independent $\phi_\perp$ and $h_e$, for isotropic interactions $\phi_{||} = \phi_\perp$ as for fcc(100) homoepitaxy, $L_k$ is tied to the value of $\phi_\perp$ ($= \phi_{||}$).

Rates $h_\pm$ for attachment to and detachment from straight steps have been described in the introduction to Sec.2. For completeness, we note that while direct attachment to kinks from terraces occurs at rate $h_\pm = \exp(-\beta\delta_\pm)h$, detachment from kinks directly to terraces occurs at rate $\exp(-\beta\phi_\perp -\beta\phi_{||} -\beta\delta_\pm)h$. Also, direct attachment of an edge adatom to a kink occurs at rate $h_e$ (as we do not include a separate kink attachment barrier along steps), and detachment from the kink to an edge site occurs at rate $\exp(-\beta\phi_{||})h_e$. See Refs. [18,21] for further discussion.

For schematics of the 2D steady-state density profiles, n(i, j), from solving the 2D DDE, we refer the reader to Ref.[18,21]. Naturally, these tend to have a global maxima in the center of the terraces furthest away from the kink sites which act as sinks for the excess adatom density. Along the step edge, the excess adatom density has maximum in between kink sites, i.e., $\delta n(i,0) = n(i,0) - n_{eq} > 0$ is maximized in between kink sites.

For extraction of $K_\pm$ and P, our strategy is again to make a connection with the quasi-1D continuum formulation of Sec.2A. To this end, it is appropriate to consider the average along the step of the density profile, n(i,j), in the discrete 2D DDE model. This average operation has the form $<n(j)> = (1/l_k) \sum_{0 \leq i < l_k} n(i,j)$. In perhaps the most natural formulation, one also calculates the average fluxes $<J_+>$ from row j = 1 to the step j = 0, and $<J_->$ from row j = -1 to the step j = 0. Adatom densities at the step edge $<n_\pm>$ can be identified with either $<n(\pm 1)>$ or they can be obtained by extrapolating terrace densities to the step edges denoted by $<n(0\pm)>$. The former is the non-extrapolation case N in the notation of Sec.2B, and the latter is the extrapolation case E. Then, for a uniform deposition flux F, the $K_\pm$ and P are defined to satisfy

$<J_\pm> = a^{-2} K_\pm (<n_\pm> - n_{eq}) + a^{-2} P(<n_\pm> - <n_\mp>)$.           (14)

Since both $<J_\pm>$ and averaged densities are directly proportional to F, the $K_\pm$ and P are independent of F. We should emphasize that there is additional flexibility in the identification of $<J_\pm>$. An alternative to the above prescription might determine $<J_+>$ from the average flux from row j = 2 to j = 1, and $<J_->$ from row j = -2 to j = -1. This choice, together with the identification of $<n_\pm>$ as $<n(\pm 1)>$, will be denoted by M for modified flux choice. See also Appendix A for a discussion for the corresponding 1D DDE models.

Finally, we note that for the 2D DDE model, we can directly calculate the adatom density at the step edge, and thus extract the averaged quantity $<n(0)>$. As in the discussion of the 1D DDE, we claim that it is natural to consider the rescaled version of this density $<n^*(0)> = \exp(-\beta\phi_\perp)<n(0)>$ form comparison with adatom densities on the terrace. Of particular relevance for the analysis in Sec.3A-C for uniform deposition is the non-trivial result that the rescaled density at the step edge, $<n^*(0)>$, corresponds to the extrapolated density, $<n(0+)>$ [$<n(0-)>$], in the case of no additional attachment barrier $\delta_+$ [$\delta_-$]. This behavior is analogous to that discussed for the 1D DDE.

## 2E. DISCRETE 2D DDE MODEL: EXTRACTION OF $K_\pm$ AND P

Once the averaged fluxes, $<J_\pm>$, and the averaged densities, $<n_\pm>$, are determined from analysis of the 2D DDE model, (14) for uniform deposition only yields two relations for three quantities. Consequently, $K_\pm$ and P cannot be uniquely determined in this way. One exception to this scenario is when the steps have symmetric attachment barriers (or no attachment barriers), and as a consequence one has $<n_+> = <n_->$ so the permeability term is absent in (14). Then $<J_+> = <J_->$ and $K = K_+ = K_-$ is determined from the single relation (14) which becomes $<J_\pm> = a^{-2} K <\delta n_\pm>$, but here P is still undetermined [18]. Another exception is when just $h_+ = 0$ since $\delta_+ = \infty$ (infinite Ehrlich-Schwoebel barrier) and thus $K_+ = 0$ (or just $h_- = 0$ since $\delta_- = \infty$ so $K_- = 0$). Now P = 0 and again (14) reduces to a single relation for a single non-zero K [18].

To determine the individual $K_\pm$ and P in the general case, and also to determine P for symmetric barriers, an alternative analysis is required. To this end, we consider situations with <u>non-uniform deposition</u>. A default choice is to select different deposition rates $F_{ti}$, for i = 1 or 2, on alternating terraces. More precisely, the rate $F_{t1}$ ($F_{t2}$) will apply for non-step edge sites nw < j < (n+1)w for even (odd) n. For step edge rows, it is convenient to have the flexibility to separated specify deposition rates $F_{ei}$, for i = 1 or 2, where $F_{e1}$ ($F_{e2}$) applies for j = nw with even (odd) n. One could set $F_{e1} = F_{t1}$ and $F_{e2} = F_{t2}$ corresponding to uniform deposition on each terrace. See Appendix B. However, it will be more convenient to set $F_{e1} = F_{e2} = F_{t1}$ (or $F_{t2}$) as this choice ensures symmetry of the density profiles for symmetric attachment barriers (as discussed further below).

The feature that non-uniform deposition provides a natural vehicle to assess permeability is best illustrated by considering the special case for the "extreme" choice with no deposition on type 2 terraces, so $F_{t1}$ = F and $F_{t2}$ = 0. With <u>symmetric attachment barriers</u> described by a single K, the excess density on type 2 terraces is uniform by symmetry, and should adopts the value

$$\delta n_2 = n_2 - n_{eq} \approx \tfrac{1}{2}\, WFP/[K(K+2P)], \tag{15}$$

based on the continuum analysis (6). Thus, $\delta n_2$ only has significant non-zero values in the presence of permeability P>0. Asymmetric attachment barriers produce a more complicated scenario given the linear variation in adatom density across the type 2 terrace. See Sec.4.

Naturally, density profiles for any choice of $F_{t1} \neq F_{t2}$ will incorporate information on P. For the general case of $F_{t1} > F_{t2}$, we now describe the strategy to extract K's and P, but also comment on additional perhaps unanticipated complications with this analysis. In general, there is distinct behavior at ascending and descending steps on each terrace. We initially assign distinct kinetic coefficients $K_{1\pm}$ ($K_{2\pm}$) for the type 1 (type 2) terrace, where physically one would expect that $K_{1+} = K_{2+}$ and $K_{1-} = K_{2-}$ since there is only a single type of step in the model. Behavior of fluxes and densities at the step edges give <u>four relations</u> determining these K's in terms of P. We can for example use these relations to obtain $K_{1\pm} = K_{1\pm}(P)$ and $K_{2\pm} = K_{2\pm}(P)$ as functions of an unknown P. We can then demand that either the $K_+$ agree on both terraces, i.e. $K_{1+}(P) = K_{2+}(P)$ yielding P = $P_+$. However, instead one could demand that the $K_-$ agree on both terraces, i.e., $K_{1-}(P) = K_{2-}(P)$ yielding P = $P_-$. In a fully consistent theory, one would have $P_+ = P_-$, and also the K's and P from this analysis would be consistent with the relationships determined from the density profiles. However, while $P_+$ and $P_-$ are generally close, they are not exactly equal.

We emphasize, however, that it is not reasonable to expect that unique consistent $K_\pm$ and P can be extracted from the discrete 2D DDE model for arbitrary choices of parameters. Such a 2D atomistic-level models cannot be exactly described within a 1D continuum formalism. Even in simple cases where P is not relevant, $K_\pm$ values depend upon the interpretation of the adatom density at step edges. In contrast to the traditional continuum picture, we also know that $K_\pm$ depend on numerous details of the system such as terrace widths and in fact on the entire terrace distribution [18].

One exception avoiding the above inconsistency is the case with <u>symmetric attachment barriers</u> and where $F_{e1} = F_{e2} = F_{t1}$ or $F_{t2}$. Then, by symmetry of the adatom profile about the center of the terraces, one has that $K_{1+}(P) = K_{1-}(P)$ and $K_{2+}(P) = K_{2-}(P)$ and consequently that $P = P_+ = P_-$ is uniquely determined. Notwithstanding, we still find slightly different values of K's and P depending on whether we select $F_{e1} = F_{e2} = F_{t1}$ or $F_{e1} = F_{e2} = F_{t2}$. However, one can argue that it is most appropriate to utilize limiting values as $F_{t2}/F_{t1} \to 1$ corresponding to physical uniform deposition where the discrepancy disappears. In the asymmetric case, a slight difference persists in K's and P's even after taking this limit.

## 3. PERMEABILITY AND KINETIC COEFFICIENTS FOR SYMMETRIC ATTACHMENT

In this case, one has that $K_+ = K_- = K$. Our goal is to determine not just this single K, but also P. Again, we use E [N] to denote the case where $\rho_\pm$ are interpreted as the extrapolated $<n(0\pm)>$ [as the non-extrapolated $<n(\pm 1)>$], and where fluxes $<J_\pm>$ are from rows j = ±1 to the step edge. M denotes a modified treatment where $<J_\pm>$ are from rows j = ±2 to j = ±1 and $\rho_\pm$ are interpreted as $<n(\pm 1)>$. Various simple relationships between $K_\pm$ for these different formulations are described in Appendix C. Again, for uniform deposition, we recall that $K = a^2 <J_\pm>/<\delta n_\pm>$ where $<J_+> = <J_->$ and $<\delta n_+> = <\delta n_->$.

### 3A. BASIC BEHAVIOR FOR ZERO ATTACHMENT BARRIERS

Our default analysis will involve steps with substantial kink separation $L_k = 24a$ corresponding to $\beta\phi_\parallel = 7.66$. We also set $\phi_\perp = \phi_\parallel$ and choose "slow" edge diffusion with rate $h_e = \exp(-\beta\phi_\perp)h$ (so that results are independent of $\phi_\perp$) and terrace width W = 40a. We consider deposition with differing rates $F_{t1} > F_{t2}$ on alternating terraces. For the extreme case of $F_{t2} = 0$, the averaged adatom density profile is shown in Fig.5, and one obtains for the non-extrapolation approach (in units of ah)

$$P(N) = 0.47980 \text{ (or 0.47881) and } K(N) = 0.04040 \text{ (or 0.04238)}, \tag{16}$$

setting step edge deposition rates as $F_{e1} = F_{e2} = F_1$ (or $F_{e1} = F_{e2} = F_2$). The high value of P and low value of K (despite the lack of step attachment barriers) reflects the large kink separation which inhibits incorporation at kink sites.

The discrepancy in values of P and K for different choices of edge deposition is eliminated by considering behavior as $F_{t2}/F_{t1} \to 1$, as shown in Fig.6. This yields the unique limiting values $P(N) = 0.47932$ and $K(N) = 0.04137$, differing only slightly from the values for $F_{t2}/F_{t1} = 0$. A similar scenario applies using the modified (M) approach where distinct limiting values of $P(M) = 0.45474$ and $K(M) = 0.03925$ are found. Significantly, all of these limiting $K_\pm$ equal the ones found in a standard analysis [18] for uniform deposition ($F_{e1} = F_{e2} = F_{t2} = F_{t1}$), from which P cannot be determined, so our methods to extract the $K_\pm$ are consistent.

For the extrapolation (E) approach, one finds that $K(E) = 0.04315$ and $P(E) = \infty$, the latter result reflecting the feature that $<n(0+)> = <n(0-)>$, so that $\rho_+ = \rho_-$ and any discrepancy between $J_\pm$ and $J_{K\pm}$ forces infinite P. The feature that $P = \infty$ might also be anticipated from our discrete 1D DDE analysis for case E in Sec.2C. Thus, in Sec.3B-C for zero attachment barrier, we do not report values for infinite P(E).

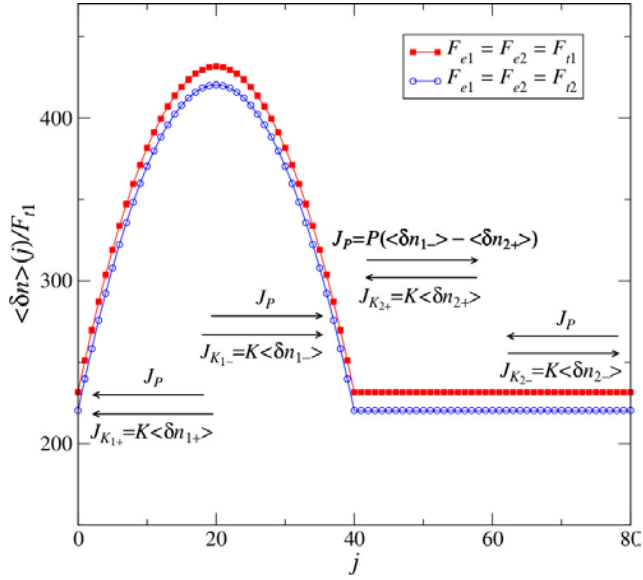

Fig.5. Average adatom density profile $\langle\delta n\rangle$ rescaled by flux $F_{t1}$ as a function of terrace position j. Data for $W = 40a$, $L_k = 24a$, $\delta_- = \delta_+ = 0$ and deposition fluxes $F_{t1} > 0$ and $F_{t2} = 0$. The arrows pointing right (left) indicate the diffusive fluxes for descending (ascending) steps, at j=0, 40 and 80. These show the cancellation of flux on the right terrace 2.

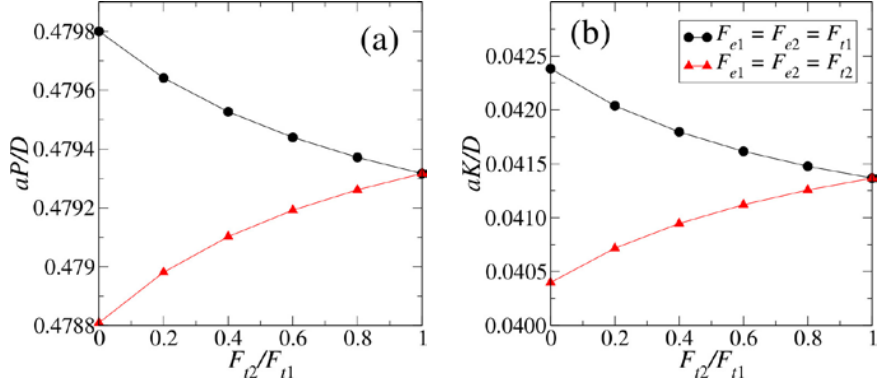

Fig.6. Dependence of (a) P(N) and (b) K(N) (from different choices of deposition fluxes at step edges) on ratio $F_{t2}/F_{t1}$. These data are for approach N with $W = 40a$, $L_k = 24a$, $\delta_- = \delta_+ = 0$, but similar behavior is found for other approaches and parameters.

### 3B. DEPENDENCE ON KINK SEPARATION FOR ZERO ATTACHMENT BARRIERS

Recalling the relation, $L_k \approx \frac{1}{2} a \exp(\frac{1}{2}\beta\phi_\parallel)$ [2], for kink separation, we adjust $L_k$ by adjusting $\beta\phi_\parallel$. First, let us consider $\phi_\perp = \phi_\parallel$, as in Sec. 3A, retaining an edge diffusion rate $h_e = \exp(-\beta\phi_\perp)h$ so that behavior is independent of $h_e$ and $\phi_\perp$. In this analysis, we set $W = 60a$. From previous analyses with uniform deposition for zero or symmetric attachment barriers, it was shown that K naturally decreases with increasing kink separation, $L_k$ [18]. Here, we extend this analysis using non-uniform deposition to also assess behavior of P.

Results in the limit as $F_{t2}/F_{t1} \to 1$ for P and K versus $L_k$ are shown in Fig.7. For cases N and M, one finds that P increases with $L_k$ but saturates. Values depend on the choice of approach, the slight difference actually increasing with $L_k$. Presumably, an expected increase of P with increasing $L_k$ is counterbalanced by the effect of increasing $\phi_\perp$ to produce saturation. We also find that $K \sim ah/(L_k/a)^2$, as $L_k \to \infty$, for any of the approaches N, M, or E [18]. This result can be understood from a rough analysis noting that $<J_\pm> \approx \frac{1}{2} a^{-2} F W$, and that the rescaled excess adatom density at the step edge follows from analysis of a 1D deposition-diffusion equation with adatoms impinging at rate $<J_\pm>$ and diffusing with hop rate h to sinks separated by $L_k$. Thus, analysis of this 1D problem yields $<\delta n*(0)> = <\delta n_\pm(E)> \sim <J_\pm>(L_k)^2/(ah)$ recovering the above form for K. Behavior of $<\delta n_\pm(E)>$ and $<\delta n_\pm(N)>$ should be similar in this case. A more complex semi-continuous version of this analysis can be found in an Appendix of Ref. [18]. The data shown in Fig.7 for smaller $L_k$ does not show pure asymptotic $1/(L_k)^2$ scaling. However, behavior in this regime can be reasonably described by the more general form

$$K \sim ah/[(L_k/a)^2 + B(L_k/a) + C]. \tag{17}$$

The decrease in K with increasing $L_k$ is certainly expected as capture at far-separated steps is inhibited.

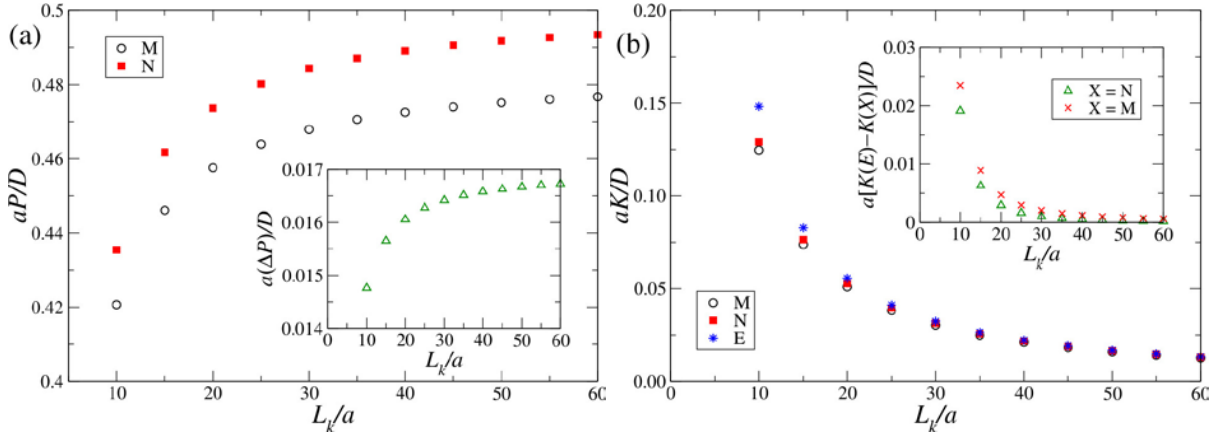

Fig.7. Variation of (a) P's and (b) K's (from different approaches M, N and E) with the kink separation $L_k$. The insets show the differences $\Delta P = P(N)-P(M)$, $K(E)-K(N)$, and $K(E)-K(M)$, versus $L_k$. Data for $W = 60a$, $\delta_- = \delta_+ = 0$ and $F_{t2}/F_{t1} \to 1$.

### 3C. DEPENDENCE ON $h_e$ AND $\beta\phi_\perp$ FOR ZERO ATTACHMENT BARRIERS

As already noted, defining $h_e = \exp(-\beta\phi_\perp)h$ makes P and K independent of $\phi_\perp$, since only the combination $\exp(\beta\phi_\perp)h_e$ appears in the rescaled 2D DDEs. However, in this section, we explore the more general case where $h_e$ and $\phi_\perp$ are regarded as independent parameters. For uniform deposition, behavior of K as a function of $h_e$ can be assessed from the relation, $K = a^2<J_\pm>/<\delta n_\pm>$. For this case where edge diffusivity is decoupled from binding to the step edge, one expects that the rescaled excess adatom density right at the step edge satisfies $<\delta n*(0)> \sim \exp(-\beta\phi_\perp)h/h_e \times (a<J_\pm>/h)$, for large $h_e/h$ or $\beta\phi_\perp$. This result reflects the feature that $\langle\delta n*(0)\rangle$ should scale inversely with $h_e$ based on a simplified 1D analysis. The first factor in $\langle\delta n*(0)\rangle$ reduces to unity for the choice $h_e = \exp(-\beta\phi_\perp)h$, thus recovering standard results for that case. As noted above, for this case with zero attachment one has that $\langle\delta n_\pm(E)\rangle = \langle\delta n*(0)\rangle$. This result, together with a relation in Appendix C allowing assessment of $\langle\delta n_\pm(N)\rangle$, yields

$$K(E) \sim ah_e \exp(\beta\phi_\perp), \text{ and } K(N) \sim ah_e \exp(\beta\phi_\perp)/[c + (h_e/h)\exp(\beta\phi_\perp)], \quad (18)$$

for $c = O(1)$ so that $K(N) \sim ah$ for large $h_e/h$ or $\beta\phi_\perp$. This behavior is confirmed by results in Fig.8b for fixed $\phi_\perp$ and different ratios $h_e/h$, and also in Fig. 9b for varying $\phi_\perp$ with $h_e = h$ kept constant. Clearly, enhanced edge diffusion enhances capture at kink sites resulting in higher values of K. Enhanced binding at step edges also naturally produces enhanced adatom capture and enhanced K (provided that $h_e$ does not decrease like $\exp(-\beta\phi_\perp)$ as $\phi_\perp$ increases).

For the analysis of permeability (for the case N) based on behavior for differing deposition fluxes on alternating terraces, it is perhaps simplest to consider the extreme case of no deposition on terrace 2. Then, the relation (15) together with the assumption that $P(N) \ll K(N)$ and $\langle J_1\rangle \approx \frac{1}{2} FW a^{-2}$ yields $P(N) \sim a^{-2} K(N)^2 \langle\delta n_2\rangle/\langle J_1\rangle$. Then, using that $K(N) \sim ah$ for large $h_e/h$ or large $\beta\phi_\perp$, and the relation $\langle\delta n_2\rangle \sim \exp(-\beta\phi_\perp)h/h_e \times (a\langle J_1\rangle/h)$ mimicking behavior noted above for $\langle\delta n*(0)\rangle$ for uniform deposition, one concludes that

$$P(N) \sim (ah) \times \exp(-\beta\phi_\perp)h/h_e, \text{ for large } h_e/h \text{ or } \beta\phi_\perp. \quad (19)$$

See Appendix C for an alternative analysis. The behavior predicted by (19) is confirmed in Fig.8a and Fig.9a. Just as enhanced edge diffusion enhances capture at kinks, it also inhibits transport across steps (without capture). Likewise, enhanced binding at step edges naturally inhibits transport across steps.

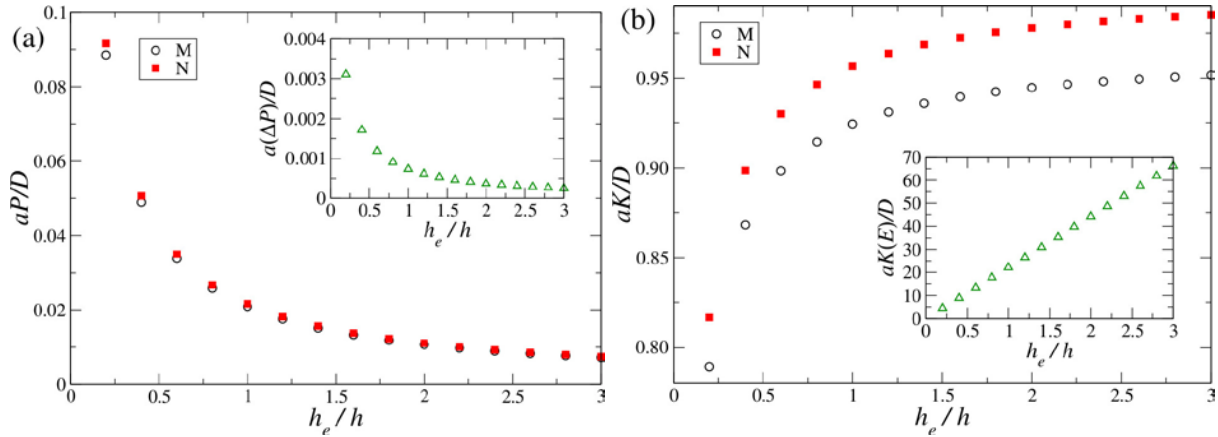

Fig.8. Main plots: dependence of (a) P's and (b) K's (for approaches M and N) with the ratio of edge on terrace hopping rates $h_e/h$. Insets: (a) difference $\Delta P=P(N)-P(M)$ and (b) K(E) versus $h_e/h$. Data for $W = 60a$, $L_k = 30a$, $\delta_- = \delta_+ = 0$ and $F_{t2}/F_{t1} \to 1$.

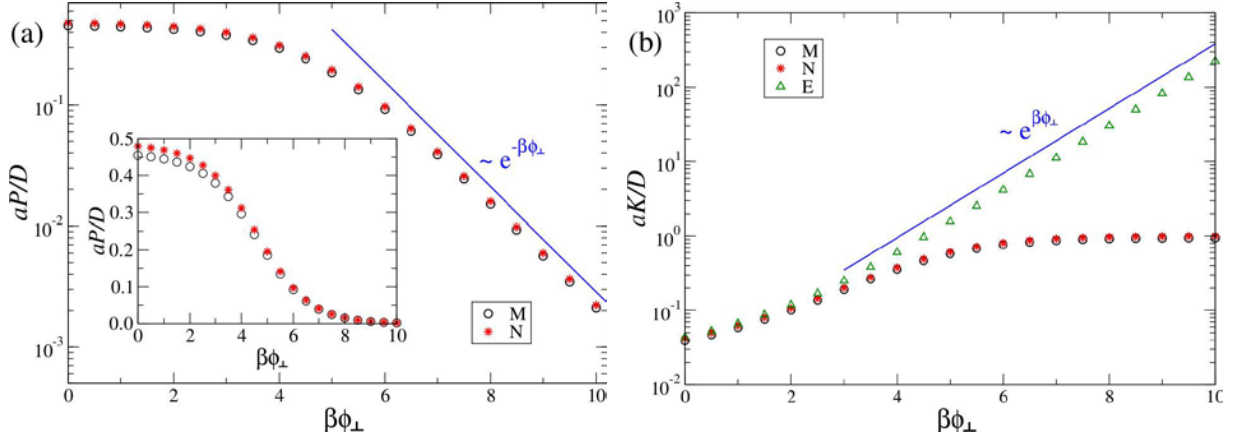

**Fig.9.** Variation of (a) P's and (b) K's (for different approaches) with NN adatom-edge interaction $\beta\phi_\perp$. In (a), the inset shows the same data of main plot in linear scaling. Data for W=40a, $L_k$ = 24a, $\delta_- = \delta_+ = 0$ and $F_{t2}/F_{t1} \to 1$.

### 3D. NON-ZERO SYMMETRIC ATTACHMENT BARRIERS

Now P(E) is finite, as well as P(N) and P(M), and we will see that these different formalisms produce similar behavior. In this case, we choose $h_\pm = \exp(-\beta\delta)h$ and again set $h_e = \exp(-\beta\phi_\perp)h$. It is intuitively clear that both K → 0 and P → 0, as $\beta\delta \to \infty$. Furthermore, for large attachment barriers, the adatom density on the terrace becomes more uniform including in the direction along the step. Thus, the discrete 1D DDE model should more accurately describe behavior in the 2D model, and the prediction from (9) or (11) that

$K_\pm \approx ah \exp(-\beta\delta)$ (20)

should apply. Indeed, the results in Fig.10 show that this behavior is realized for the 2D model after a crossover from non-asymptotic regime for small $\beta\delta$. To elucidate the behavior of P, again the discrete 1D DDE model provides insights noting that the relaxation rate describing incorporation of adatoms which have already reached the step edge will not decrease with increasing $\beta\delta$ as such adatoms have already surmounted the step attachment barrier and just need to diffuse along the step edge to reach kink sites. Thus, the result from (10) and (12) that P ≈ $ah^2\exp(-2\beta\delta)$/r should be applicable. Even though a precise expression for r is not available, this quantity will not depend on $\beta\delta$. Results in Fig.10 confirm the variation

P ~ ah exp(-2$\beta\delta$). (21)

It is also appropriate to note that the behavior of K and P is not sensitive to the detailed prescription (extrapolated, non-extrapolated or modified) of these quantities.

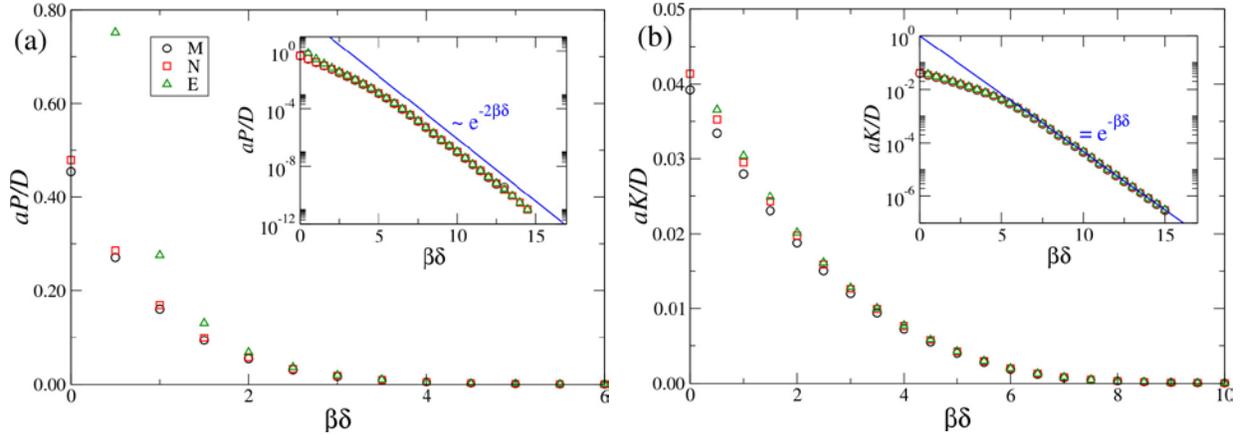

Fig.10. (a) Permeability and (b) kinetic coefficient versus interaction $\beta\delta$, with $\delta = \delta_+ = \delta_-$ in linear (main plot) and log-linear (inset) scales. Data for different approaches for $W = 40a$, $L_k = 24a$, and $F_{t2}/F_{t1} \to 1$ are shown.

## 4. PERMEABILITY AND KINETIC COEFFICIENTS FOR ASYMMETRIC ATTACHMENT

Again in analyzing K's and P, case E (case N) indicates that $\rho_\pm$ are obtained from extrapolated $<n(\pm0)>$ (non-extrapolated $<n(\pm1)>$), and fluxes $<J_\pm>$ are from rows $j = \pm 1$ to the step edge. M indicates that $<J_\pm>$ are from $j = \pm 2$ to $j = \pm 1$ and $\rho_\pm$ are interpreted as $<n(\pm1)>$. Our default analysis will involve steps with substantial kink separation $L_k = 24a$ corresponding to $\beta\phi_{||} = 7.66$, $h_e = \exp(-\beta\phi_\perp)h$, and terrace width $W = 40a$.

## 4A. BASIC BEHAVIOR FOR NON-ZERO ES BARRIER ($\delta_- >0$, $\delta_+ =0$)

For the case $\delta_- > 0$ (corresponding to a non-zero Ehrlich-Schwoebel (ES) barrier for downward transport at step edges) and $\delta_+ = 0$ (facile attachment at ascending steps), it is clear that $K_- < K_+$. We consider deposition with differing rates $F_{t1} > F_{t2}$ on alternating terraces. For the extreme case of non-uniform deposition with $F_{t2} = 0$, the averaged adatom density profile is shown in Fig.11. Again the non-zero excess density of terrace 2 reflects the presence of a non-zero permeability $P > 0$, and the linear variation across the terrace is a simple consequence of the lack of deposition which implies a constant diffusion flux across the terrace. Direction of the net flux across terrace 2 is perhaps not obvious as it is in the direction of the smaller of the two permeability fluxes impinging on terrace 2. This behavior also relies on the feature that $K_- << K_+$.

Next, we present results for $K_\pm$ and P's based on an analysis as described in Sec. 2E for the limiting case of quasi-uniform deposition ($F_{t2}/F_{t1} \to 1$). Specifically, after determining the relations $K_{1\pm}(P)$ and $K_{2\pm}(P)$ from analysis of fluxes to step edges and excess adatom densities on both terraces, we determine $P_+$ and $P_-$ from the relations $K_{1+}(P_+) = K_{2+}(P_+)$ and $K_{1-}(P_-) = K_{2-}(P_-)$. For each of the approaches E, N, and M, one finds small differences in $P_+$ and $P_-$. See Table I for some examples. For cases N and M, $K_-$ is about 40% of $K_+$ for $\beta\delta_- = 0.8$, and about 3% of $K_+$ for $\beta\delta_- = 3.5$. P values are significantly higher than $K_+$ values for $\beta\delta_- = 0.8$, but about 50% lower for $\beta\delta_- = 3.5$. For case E, one finds somewhat higher values for $K_+$ and P, but extremely small values for $K_-$.

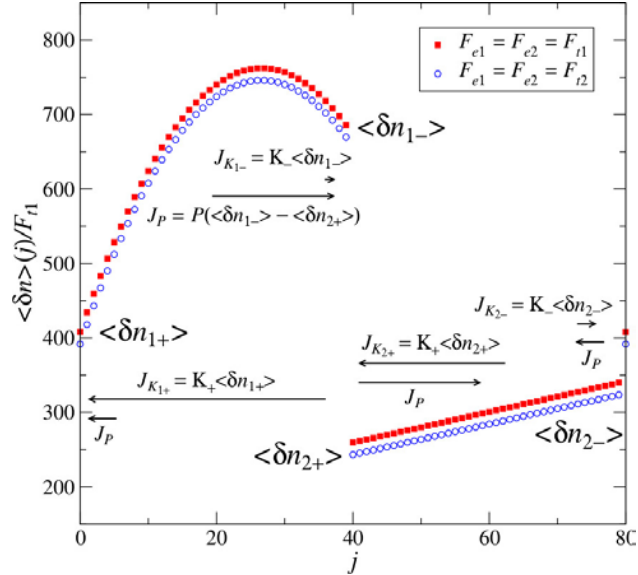

Fig.11. Average adatom density profile <δn> rescaled by flux $F_{t1}$ as a function of terrace position j, for W = 40a, $L_k$ = 24a, $\delta_+ = 0$, $\delta_- = 3.5$ and deposition fluxes $F_{t1} > 0$ and $F_{t2} = 0$. The arrows pointing right (left) indicate the diffusive fluxes for descending (ascending) steps, located at j=0, 40 and 80. Note that the flux is constant across the right terrace 2.

| $\beta\delta_-$ | Approach | $aP_+/D$ | $aP_-/D$ | $a(P_+-P_-)/D$ | $aK_+/D$ | $aK_-/D$ |
|---|---|---|---|---|---|---|
| 0.8 | M | 0.27954 | 0.27898 | 0.00056 | 0.04813 | 0.02155 |
|  | N | 0.29446 | 0.29426 | 0.00020 | 0.05066 | 0.02276 |
|  | E | 0.81651 | 0.81597 | 0.00055 | 0.07736 | ≈ 0 |
| 3.5 | M | 0.02681 | 0.02607 | 0.00075 | 0.05303 | 0.00120 |
|  | N | 0.02805 | 0.02769 | 0.00036 | 0.05510 | 0.00166 |
|  | E | 0.03155 | 0.03114 | 0.00041 | 0.06010 | ≈ 0 |

Table I. Permeabilities $P_\pm$ and kinetic coefficients $K_\pm = K_{1\pm}(P_\pm) = K_{2\pm}(P_\pm)$ for approaches M, N and E. The slight difference between $P_+$ and $P_-$ is shown in 5$^{th}$ column. Data are obtained for W=40a, $L_k$ = 24a, $\delta_+$ = 0, and $F_{e1} = F_{e2} = F_{t1}$ with $F_{t2}/F_{t1} \to 1$.

A more complete description of how $P_+$ depends on $\beta\delta_-$ is given in Fig.12a. As might be anticipated from our discrete 1D DDE analysis, one finds decay like $P_\pm \sim$ (ah) exp(-$\beta\delta_-$) for a broad range of large $\beta\delta_-$. The variation of the difference between $P_+$ and $P_-$ depends on the approach E, N, or M, but it is always very small and tends to decrease for large $\beta\delta_-$. See Fig.12b. Behavior of $P_+$ for very large $\beta\delta_-$ has an unusual saturation feature which we will not discuss in detail here. Finally, we describe results for the corresponding behavior of $K_{\pm}$. As expected from the discrete 1D DDE analysis, one finds that $K_- \sim (ah)\exp(-\beta\delta_-)$, while $K_+$ saturates for large $\beta\delta_-$. See Fig.13.

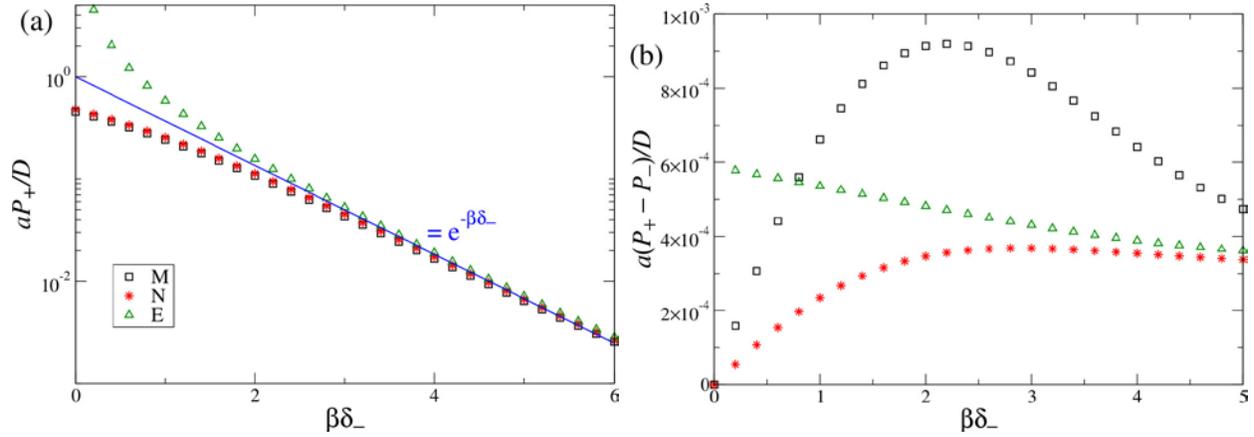

Fig.12. (a) Permeability $P_+$ and (b) differences $P_+ - P_-$, for approaches M, N and E, as functions of the ES barrier $\beta\delta_-$. Data for the parameters $W = 40a$, $L_k = 24a$, $\delta_+ = 0$ and $F_{t2}/F_{t1} \to 1$.

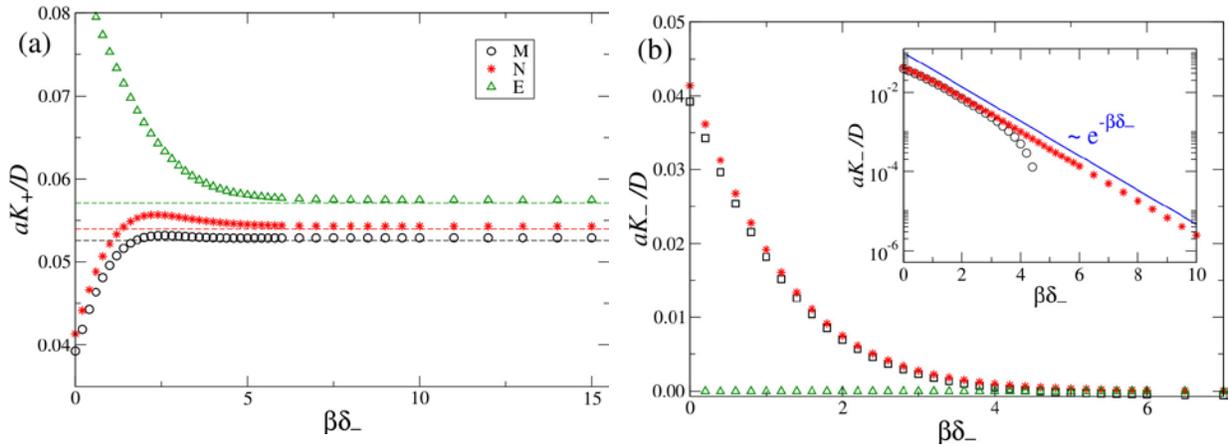

Fig.13. Kinetic coefficients (a) $K_+$ and (b) $K_-$ versus $\beta\delta_-$. In (b) the inset shows the same data of the main plot in log-linear scale. In (a), horizontal lines indicate values obtained for standard calculation [18] considering an infinity ES barrier. Parameters are the same as in Fig.12.

## 4B. BEHAVIOR FOR ASYMMETRIC ATTACHMENT BARRIERS WITH $\delta_- = 2\delta_+ = 2\delta$

Again, as in Sec.3D, one expects that, for large barriers, the discrete 1D DDE model should more accurately describe behavior in the 2D model. Thus, (9) or (11) for $K_\pm$ and (10) or (12) for P suggest

$K_\pm \approx ah \exp(-\beta\delta_\pm)$, and $P \sim ah \exp(-\beta\delta_+ -\beta\delta_-) = ah \exp(-3\beta\delta)$. (22)

Results shown in Fig.14 confirm these predictions. Note that the behavior of K and P is not sensitive to the detailed prescription (extrapolated, non-extrapolated or modified) of these quantities.

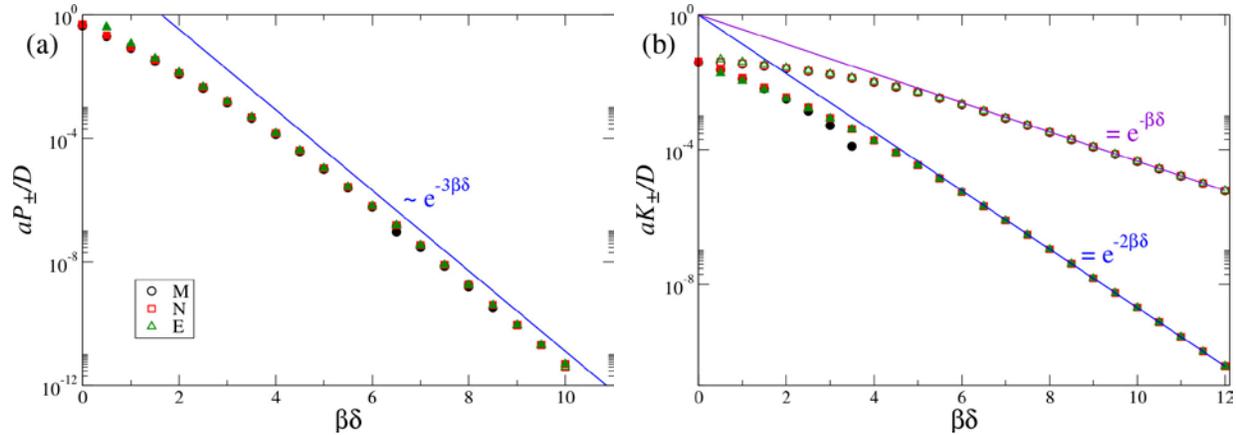

Fig.14. (a) Permeabilities $P_\pm$ and (b) kinetic coefficients $K_\pm$, for approaches M, N and E, as functions of $\beta\delta$, with $\delta_+ = \delta$ and $\delta_- = 2\delta$. Open (full) symbols represent quantities + (−). Data obtained for $W = 40a$, $L_k = 24a$, and $F_{t2}/F_{t1} \to 1$.

## 5. DISCUSSION AND CONCLUSIONS

Our analysis shows that the discrete 2D DDE formulation is particularly effective at not just elucidating the general behavior of permeability, P, and kinetic coefficients, $K_\pm$, but also in quantifying these parameters. The latter is necessary for application to the description of specific systems where appropriate energetic and geometric parameters would provide input to our 2D DDE formulation. Steady-state analysis of non-uniform deposition scenarios allows determination of each of P and $K_\pm$. This is not possible just considering uniform deposition. For the extreme case where there is no deposition on alternating terraces, one gains immediate insight into the extent of permeability from the non-zero excess adatom density on those terraces.

To conclude, we review in more detail our results for the dependence of P and $K_\pm$ on key parameters, and also discuss how these results relate to behavior in specific systems:

(i) <u>Dependence on kink separation $L_k$</u>. The behavior $K_\pm \sim a^3 h/(L_k)^2$ from (17) illustrating the decrease of $K_\pm$ with increasing $L_k$ which is expected since incorporation at kinks is naturally inhibited. This dependence is key to understanding behavior during step flow on dimer-row reconstructed vicinal Si(100) or Ge(100) surfaces which exhibits alternating rough (or meandering) steps and smooth (or stiff steps) [18,21,27,28]. Rough (smooth) steps have low (high) $L_k$ values, and thus high (low) $K_\pm$ values. Another class of systems are fcc(110) metal surfaces [29,30]. Here, steps along the <110> direction (parallel to rows of neighboring surface atoms) are smooth and stiff with large $L_k$ and low $K_\pm$, but steps in the orthogonal <001> direction are rough with low $L_k$ and higher $K_\pm$.

Dependence of P on $L_k$ depends on model details. If edge diffusivity decreases as $L_k$ increases, then P can saturate as shown in Sec.3B. However, increasing $L_k$ with other parameters fixed would naturally lead to a decrease in P as suggested by (10) or (12) using $r \sim \exp(+\beta\phi_\perp) h_e / (l_k)^2$.

(ii) <u>Dependence on edge diffusivity</u> $h_e$. The behavior $K_\pm \sim a h_e$ from (18), and $P \sim a h^2 / h_e$ from (19), reflect the feature that facile edge diffusion enhances incorporation at kinks sites and thus inhibits step crossing without capture. From this perspective, one expects steps on fcc(100) metal surfaces to have high $K_\pm$ and low P since $h_e$ far exceeds h; on the other hand, $K_\pm$ may be lower and P higher for fcc(111) metal surfaces where generally $h_e$ is far below h [5].

(iii) <u>Dependence on binding to the step edge</u> $\beta\phi_\perp$. Generally, one expects binding to the step edge to enhance capture at kinks and thus enhance $K_\pm$. However, if step edge diffusivity is reduced with increased binding, as with the choice $h_e = \exp(-\beta\phi_\perp) h$, then $K_\pm$ and P and independent of $\phi_\perp$ This "bond-breaking" specification is generally regarded as being more realistic for semiconductor rather than metallic surfaces [5].

(iv) <u>Dependence on step attachment barriers</u>, $\delta_\pm$. The traditional 1D DDE analysis (which assumes instantaneous incorporation at steps and set $r = \infty$) successfully predict the basic behavior $K_\pm \sim (ah) \exp(-\beta\delta_\pm)$, as confirmed by (9) and (11) even for finite r. This behavior is also validated by our 2D DDE analysis as given by (20) and (22). This behavior reflects the traditional expectation that such additional attachment barriers should inhibit capture at step edges and thus reduce $K_\pm$. Our refined 1D DDE analysis (10) and (12) indicated the behavior of $P \approx ah \exp(-\beta\delta_+ - \beta\delta_-)$, which was confirmed by our 2D DDE analysis. See (21) and (22). These results can be directly applied to metal surfaces where one generally finds no barrier for attachment to ascending steps ($\delta_+ = 0$) but a non-zero Ehrlich-Schwoebel barrier for attachment to descending steps ($\delta_- > 0$). Generally, $\delta_-$ is smaller for fcc(100) surfaces which would enhance P, but this is counterbalanced by a high $h_e$. On the other hand, $\delta_-$ is generally higher for fcc(111) surfaces which would reduce P, but a compensating feature is that $h_e$ is lower. For semiconductor surfaces, it is sometimes suggested that there is a barrier for attachment to ascending steps [28], which would imply a reduced $K_+$.

## ACKNOWLEDGEMENTS

RZ and JWE were supported for this work by the U.S. Department of Energy (USDOE), Office of Basic Energy Sciences, Division of Chemical Sciences, Geosciences, and Biosciences through the Ames Laboratory Chemical Physics program. The work was performed at Ames Laboratory which is operated for the USDOE by Iowa State University under Contract No. DE-AC02-07CH11358. TJO acknowledges the support from CNPq and FAPEMIG (Brazilian agencies).## APPENDIX A: DISCRETE 1D DDE FORMULATIONS

The complete set of DDE for the 1D model of Sec.2B are as follows:

$$d/dt\, n(j) = F + h[n(j+1) - 2n(j) + n(j-1)] \approx 0, \text{ for } j > 1, \quad (A1)$$

$$d/dt\, n(1) = F + h[n(2) - n(1)] + h_+[\exp(-\beta\phi_\perp) n(0) - n(1)] \approx 0, \quad (A2)$$

$d/dt\ n(0) = F + h_+[n(1) - \exp(-\beta\phi_\perp)n(0)] + h_-[n(-1) - \exp(-\beta\phi_\perp)n(0)]$

$\quad - \nu[n(0) - n_{eq}(0)] \approx 0,$  (A3)

with analogous equations for n(j<0). The structure of the above equations is simplified replacing n(0) by the rescaled density via $n^*(0) = \exp(-\beta\phi_\perp)n(0)$, introduced in Sec.2B, and using the identity $\nu[n(0) - n_{eq}(0)] = r[n^*(0) - n_{eq}]$. If necessary, we smoothly extrapolate or "analytically extend" the n(j≠0) to the values n(0±) at the step edge or step via the defining relations

$d/dt\ n(\pm1) \equiv F + h[n(\pm2) - 2n(\pm1) + n(0\pm)] \approx 0.$  (A4)

Exact expressions for the flux,

$J_+ = a^{-1} h_+[n(1) - \exp(-\beta\phi_\perp)n(0)]$ between sites j = 1 and j = 0, and  (A5)

$J_- = a^{-1} h_-[n(-1) - \exp(-\beta\phi_\perp)n(0)]$ between sites j = -1 and j=0,  (A6)

are rewritten using the above steady-state relations to have the form (4). One then extracts the expressions for $K_\pm$ and P given in (9) and (10) associating $\rho_\pm$ with non-extrapolated (N) $a^{-2}\ n(\pm1)$, or in (11) and (12) associating $\rho_\pm$ with extrapolated (E) $a^{-2}\ n(0\pm)$.

As an aside, we note that instead that the modified (M) approach identifying fluxes, $J_\pm$, as those near to rather than at the steps, e.g., choosing $J_+ = a^{-1} h[n(2) - n(1)]$ and $J_- = a^{-1} h[n(-2) - n(-1)]$, would produce slightly different expressions for $K_\pm$ and P from those given in Sec. 2B.

We have also considered a modified discrete 1D DDE model where we let r→∞ (or ν→∞) so that $n(0) = n_{eq}(0) = \exp(-\beta\phi_\parallel)$ is no longer a free variable, but we also include direct hopping between sites j = -1 and j = +1 at rate $h_p$. In analysis of this model, if we identify $J_\pm$ as fluxes right at the step, then one obtains

$J_+ = a^{-1} h_+[n(1) - n_{eq}] + a^{-1} h_p[n(1) - n(-1)]$, and  (A7)

$J_- = a^{-1} h_-[n(-1) - n_{eq}] + a^{-1} h_p[n(-1) - n(1)]$,  (A8)

For the case where $\rho_\pm$ are identified as $a^{-2}\ n(\pm1)$, (A7) and (A8) already have exactly the form of (4), immediately yielding

$K_\pm = ah_\pm$ so $\Gamma_\pm = ah/h_\pm$, and $P = ah_p$ so $\Gamma_p = ah/h_p$,  (A9)

as reported previously [18]. For the case where $\rho_\pm$ are identified as the analytically extended $a^{-2}\ n(0\pm)$, one obtains

$K_\pm = ah[h_\pm(h - h_\mp) - h_p(h_+ + h_-)] / [(h - h_+)(h - h_-) - h_p(2h - h_+ - h_-)]$, and  (A10)

$P = ah^2 h_p / [(h - h_+)(h - h_-) - h_p(2h - h_+ - h_-)].$  (A11)

Thus, (A10) and (A11) show that both $K_\pm$ and P diverge when $h_\pm = h$, and these results reduce to the expected $K_\pm = ah/(h/h_\pm -1)$ and $P = 0$ when $h_p = 0$. Choosing $J_+ = a^{-1} h[n(2) – n(1)]$ and $J_- = a^{-1} h[n(-2) – n(-1)]$ would produce slightly different expressions for $K_\pm$ and P for either treatment.

## APPENDIX B: SYMMETRY-BREAKING STEP EDGE DEPOSITION FOR $\delta_+ = \delta_-$

In a continuum model with different deposition rates, $\mathcal{F}_{t1}$ and $\mathcal{F}_{t2}$, on alternating terraces for a perfect vicinal surface with symmetric attachment barriers, one naturally finds reflection symmetry about the center of each terrace in the adatom density averaged along the step edge. See Fig.2. However, in the discrete 2D DDE model, reasonably choosing deposition rates at step edge rows as $F_{e1} = F_{t1}$ and $F_{e2} = F_{t2}$ (corresponding to uniform deposition on each terrace) actually breaks the above reflection symmetry. This is most clear in the extreme case of no deposition on terrace 2 where $F_{e2} = F_{t2} = 0$ from examination of the averaged adatom density profile across terrace 2 which would be constant in the presence of reflection symmetry. However, results for this case, shown in Fig.15 where we focus on terrace 2, show a symmetry-breaking linear variation in averaged density across the terrace. The density is naturally bounded above (below) by the constant values obtained by symmetry-preserving choices $F_{e1} = F_{e2} = F_{t1} > 0$ ($F_{e1} = F_{e2} = F_{t2} = 0$) which were utilized in Sec.2A.

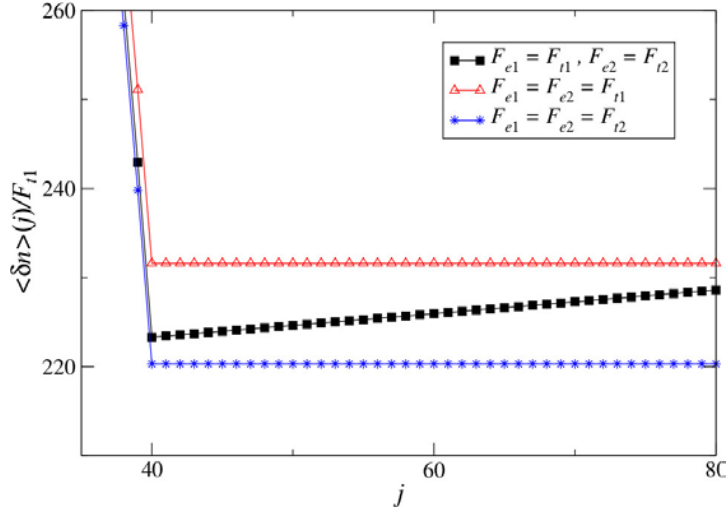

Fig.15. Average adatom density profile $<\delta n>$ rescaled by flux $F_{t1}$ as a function of terrace position j, for $W = 40a$, $L_k = 24a$, $\delta_-=\delta_+=0$ and deposition fluxes $F_{t1}>0$ and $F_{t2}=0$.

The feature that reflection symmetry is broken means that one must implement the general strategy to obtains K's and P accounting for distinct behavior at ascending and descending step edges at both terraces. Indeed, in Sec. 3 (for symmetric attachment barriers) one finds the same unique values for $K_\pm$ and P in the limit of $F_{t2}/F_{t1} \rightarrow 1$. In contrast, considering $F_{e1} = F_{t1}$ and $F_{e2} = F_{t2}$ similar but distinct $P_+$ and $P_-$ are obtained, even for $F_{t2}/F_{t1} \rightarrow 1$. See case $\delta_+ = \delta_- = 0$ in Table II. For asymmetric attachment barriers, $\delta_+ = 0$ and $\delta_- > 0$, slightly different $P_+$ and $P_-$ are found independent of the choice of fluxes at the step edges. This happens because the symmetry is naturally broken by the barriers. Anyway, comparing the values of $P_+$ and $P_-$ in Table I (for $F_{e1} = F_{e2} = F_{t1} > 0$) and in Table II one see that they are close, but in the former case the difference $P_+ - P_-$ is in general smaller. Thus, the strategy of considering identical fluxes at step edges yields improved data also for asymmetric barriers.

| $\beta\delta_-$ | Ext. appr. | $aP_+/D$ | $aP_-/D$ | $a(P_+-P_-)/D$ | $aK_+/D$ | $aK_-/D$ |
|---|---|---|---|---|---|---|
| 0.0 | M | 0.44286 | 0.46661 | -0.02374 | 0.03925 | 0.03925 |
| | N | 0.46717 | 0.49146 | -0.02429 | 0.04137 | 0.04137 |
| | E | ∞ | ∞ | -- | 0.04315 | 0.04315 |
| 0.8 | M | 0.26975 | 0.28369 | -0.01393 | 0.04772 | 0.02136 |
| | N | 0.28437 | 0.29898 | -0.01460 | 0.05024 | 0.02257 |
| | E | 0.77611 | 0.81597 | -0.03986 | 0.07550 | ≈ 0 |
| 3.5 | M | 0.02537 | 0.02616 | -0.00079 | 0.05233 | 0.00117 |
| | N | 0.02657 | 0.02775 | -0.00118 | 0.05438 | 0.00164 |
| | E | 0.02981 | 0.03114 | -0.00133 | 0.05917 | ≈ 0 |

Table II. Permeabilities $P_\pm$ and kinetic coefficients $K_\pm = K_{1\pm}(P_\pm) = K_{2\pm}(P_\pm)$ for approaches M, N and E. The difference between $P_+$ and $P_-$ is shown in 5$^{th}$ column. Data obtained for $W = 40a$, $L_k=24a$, $\delta_+ = 0$, and $F_{e1} = F_{t1}$ and $F_{e2} = F_{t2}$ with $F_{t2}/F_{t1} \to 1$.

## APPENDIX C: RELATIONSHIPS BETWEEN $K_\pm$ FOR DIFFERENT FORMALISMS

We first describe various relationships between the K's for these different approaches which follow most directly from analysis for uniform deposition at rate F per site. For symmetric attachment barriers where $<n_+> = <n_->$, $K = K_+ = K_-$ is determined from $K = a^2<J_\pm>/<\delta n_\pm>$ for excess step edge densities $<\delta n_\pm> = <n_\pm> - n_{eq}$.

Note that K(N) and K(M) share the same $<\delta n_\pm>$. Also, by symmetry, $<J_\pm(N)>$ for the same terrace must equally share the total deposition flux on that terrace except for atoms deposited directly at the step edge row. Given our specific definition of $<J_\pm(N)>$, this total flux is associated with atoms deposited on w-1 (w-3) rows for N (M), so that

$$K(M)/K(N) = <J_\pm(M)>/<J_\pm(N)> = (w-3)/(w-1). \qquad (C1)$$

This relation also reflects the more general feature that K's and P depend on the terrace width, W, generally exhibiting a 1/w type approach to limiting values as w→∞ [18]. Another natural comparison comes from the feature that K(N) and K(E) share the same $<J_\pm> = h_\pm [<\delta n_\pm(N)> - <\delta n_\pm(E)>]$. Solving this relation for $<\delta n_\pm(N)>$ in terms of $<\delta n_\pm(E)>$ and substituting into $K = a^2<J_\pm>/<\delta n_\pm>$ yields

$$K(N) = h_\pm K(E)/[h_\pm + K(E)]. \qquad (C2)$$

This relation captures the feature seen in the 1D DDE analysis that K(N) remains finite as K(E) diverges in the regime of facile incorporation at step edges.

In analyzing permeability, P, we consider deposition with fluxes $F_{t1}$ and $F_{t2}$ on alternating terraces. Symmetry implies that average fluxes to both step edge on terrace 1 are equal so that $<J_{1+}> = <J_{1-}> = <J_1>$, as are average excess densities at both steps so that $<\delta n_{1+}> = <\delta n_{1-}> = <\delta n_1>$. The same applies for terrace 2. Then, using (14) for both terraces (1 and 2) to solve for P yields

$$P = a^2 [<\delta n_2><J_1> - <\delta n_1><J_2>]/[(<\delta n_1>-<\delta n_2>)(<\delta n_1>+<\delta n_2>)]. \qquad (C3)$$

As an aside, we note that if $F_{t2} = 0$ so that $<J_2> = 0$, and if $<\delta n_1> >> <\delta n_2>$, then (C3) reduces to

$$P(N) \approx a^2 <\delta n_2><J_1>/(<\delta n_1>)^2. \qquad (C4)$$

This result matches the expression given in Sec.3C, which leads to Eq. (19), by using $<J_1> \sim a^{-2} K <\delta n_1>$.

Finally, analogous to our derivation of (C1), it immediately follows from (C3) that

$$P(M)/P(N) = (w-3)/(w-1). \qquad (C5)$$